\documentclass[10pt]{article}

\usepackage{sectsty}

\sectionfont{\fontsize{12}{14}\bfseries\sffamily}

\usepackage[super,compress]{cite}

\usepackage{amssymb}
\usepackage{amsfonts}
\usepackage{amsmath}

\usepackage{graphicx}

\usepackage{bm}

\usepackage{xcolor}

\usepackage{siunitx}

\newcommand{\dd}{\mathrm d}

\newcommand{\ii}{\mathrm i}

\newcommand{\calC}{\mathcal C}

\newcommand{\calL}{\mathcal L}
\newcommand{\calM}{\mathcal M}

\newcommand{\calQ}{\mathcal Q}

\newcommand{\barL}{{\bar L}}

\usepackage{slashed}

\newcommand\TODO[1]             {}

\definecolor{garrosgreen}{rgb}{0.1, 0.4, 0.1}
\definecolor{dartmouthgreen}{rgb}{0.05, 0.5, 0.06}
\definecolor{cambridgeblue}{rgb}{0.1, 0.3, 1.0}
\definecolor{oxfordblue}{rgb}{0.05, 0.2, 0.7}

\begin{document}

{\large \textbf{\textsf{Lorentz Breaking and $\bm{SU(2)_L \times U(1)_Y}$ 
Gauge Invariance for Neutrinos}}}

{\rm U. D. Jentschura$^{a,b}$,
I. N\'andori$^{b,c,d}$,
G. Somogyi$^b$}

{\em \scriptsize ${}^a$ Missouri University of Science and Technology,
Rolla, Missouri 65409, USA} \\
{\em \scriptsize ${}^b$ MTA--DE Particle Physics Research Group,
P.O.~Box 51, H--4001 Debrecen, Hungary} \\
{\em \scriptsize ${}^c$ MTA Atomki, P.O.~Box 51, H--4001 Debrecen, Hungary}\\
{\em \scriptsize ${}^d$ University of Debrecen, P.O.~Box 105, H--4010 Debrecen, Hungary}

\begin{abstract}
Conceivable Lorentz-violating effects in the 
neutrino sector remain a research area  of great general
interest, as they touch upon the very foundations
on which the Standard Model and our general
understanding of fundamental interactions is laid. 
Here, we investigate the relation of 
Lorentz violation in the neutrino sector
in light of the fact that neutrinos and corresponding 
left-handed charged leptons form
$SU(2)_L$ doublets under the electroweak 
gauge group. Lorentz-violating effects thus 
cannot be fully separated from questions related 
to gauge invariance.
The model dependence of the effective interaction 
Lagrangians used in various recent investigations
is investigated, with a special emphasis on 
neutrino splitting, otherwise known as 
neutrino-pair Cerenkov radiation, NPCR,
and vacuum pair emission
(electron-positron-pair Cerenkov radiation, LPCR).
We investigate two scenarios in which 
Lorentz violating effects do not 
necessarily also break electroweak gauge invariance.
The first of these involves a restricted set of gauge 
transformation, a subgroup of $SU(2)_L\times U(1)_Y$, while in the 
second, differential Lorentz violation is 
exclusively introduced by the mixing of the 
neutrino flavor and mass eigenstates.
Our study culminates in a model which fully 
preserves $SU(2)_L \times U(1)_Y$ gauge 
invariance, involves flavor-dependent Lorentz-breaking
parameters, and still allows for NPCR and LPCR decays 
to proceed.
\end{abstract}


\tableofcontents

\newpage

\section{Introduction}

Recently, tight bounds on 
Lorentz-violating parameters for neutrinos
have been derived from astrophysical
observations~\cite{StSc2014,StEtAl2015,SoNaJe2019},
based on on the notion that 
neutrino decay into electron-positron
pairs (called ``lepton-pair Cerenkov radiation'', LPCR,
or ``vacuum pair emission'') 
becomes kinematically allowed
under a Lorentz noninvariance
of the neutrino dispersion relation.
One observes that even a slight violation
of Lorentz invariance at high 
energy would lead to a large deviation
of the dispersion relation from the light 
cone (``virtuality'') $E^2 - \vec p^{\,2} = 
\vec p^{\,2} (v^2 - 1)$,
where $v > 1$ is the velocity parameter,
because of the multiplicative prefactor
$\vec p^{\,2} \approx E^2$ (in front of $v^2 - 1$) 
which grows without bound at high energy.
For high energy, the quantity 
$q^2 = E^2 - \vec p^{\,2}$
exceeds the electron-positron pair 
production threshold.

Very recently, these calculations have
been supplemented by 
an analysis of the neutrino splitting process~\cite{SoNaJe2019}
($\nu \to \nu \bar\nu \nu$,
neutrino-pair Cerenkov radiation, NPCR),
which in contrast to 
charged-lepton-pair Cerenkov radiation~\cite{CoGl2011,BeLe2012}
($\nu \to \nu \, e \, e^+$)
has negligible threshold
and can serve to set even tighter
bounds on the Lorentz-violating parameters.

On the theoretical side, 
corresponding calculations are
mainly based on the notion that Lorentz
noninvariance is restricted to the 
neutrino sector, 
while the Lorentz-violating parameter
$\delta_e = v_e^2 -1 $ is set equal to 
zero for electrons and positrons.
I.e., one assumes that the maximum attainable
velocity for electrons is exactly equal to 
$v_e = c$, where $c$ is the speed of 
light~\cite{CoGl2011,BeLe2012,St2014,StSc2014,StEtAl2015}.
One might argue, though, that 
electrons and neutrinos enter an $SU(2)_L$ doublet,
so that in addition to Lorentz violation,
also gauge symmetry is violated if one
assumes a different propagation velocity
for charged leptons and neutrinos 
in the high-energy limit.

We should point out that 
Lorentz violation for charged particles can often be studied 
more easily using other kinds of processes 
(involving electrons, positrons, and photons, for example).  
The fact that the Lorentz violation coefficients for charged 
leptons and neutrinos are not independent, 
due to the existence of $SU(2)_L \times U(!1)_Y$ 
gauge symmetry, has already been used in the 
literature~\cite{NoVo2014,BeKoLi2016}. 
However, for a number of reasons, 
it has been of prime interest to infer bounds
for Lorentz-violating parameters in the neutrino sector,
even if the models involved might be considered
as ``kinematics-only'' approaches and lead to a 
slight, perturbative breaking of
gauge invariance, to the extent to be discussed below.

Two models have  been investigated in this context by 
Bezrukov and Lee~\cite{BeLe2012} in order to analyze the 
decay of superluminal neutrinos by electron-positron
pair emission, one (``model I''), in which the normal 
Lorentz metric enters the interaction Lagrangian,
and another one (``model II''), in which 
the same Lorentz-violating ``metric''
\begin{equation}
\widetilde g^{\mu\nu}(v) = 
\widetilde g_{\mu\nu}(v) = 
{\rm diag}(1, -v, -v, -v)
\,,
\qquad v > 1 \,,
\end{equation}
enters as in the dispersion relation of the 
decaying neutrino. (The ``metric'' is noncovariant, 
hence there is no distinction between upper and lower indices.) 
Potentially, this ``metric'' (or, ``pseudo-metric'') also enters the
effective interaction Lagrangian 
describing the decay process, where $v \geq 1$
is a Lorentz-violating parameter,
which can be different for the maximum velocities 
of the initial ($v= v_i$) and final ($v= v_f$) 
particles in the decay process.
(In this paper, we have $\hbar = c = \epsilon_0 = 1$.)

In Ref.~\citen{SoNaJe2019}, an even more general
approach is taken, and the metric
entering the interaction Lagrangian is taken
in the form
\begin{equation}
\widetilde g^{\mu\nu}(v_{\rm int}) =
{\rm diag}(1, -v_{\rm int}, -v_{\rm int}, -v_{\rm int}) \,.
\end{equation}
where $v_{\rm int}$ is not necessarily equal to 
$v_i$ or $v_f$.

Here, we show in detail 
how the parameters of the models used by 
Cohen and Glashow~\cite{CoGl2011}, by Bezrukov and Lee~\cite{BeLe2012},
and by us in Ref.~\citen{SoNaJe2019}, are related 
to the gauge invariance under the $SU(2)_L$
group, and how the formulation of 
the gauge sector relates to the 
individual Lorentz-violating 
parameters of the neutrino flavor and mass eigenstates,
and those of the charged leptons.

In particular, we can anticipate that the 
model used by Bezrukov and Lee~\cite{BeLe2012} for 
vacuum pair emission turns out to be gauge 
invariant only with respect to a restricted subgroup 
of the set of $SU(2)_L$ gauge transformations;
we will investigate the respective subgroup.
 
This is important for a general understanding
of the relation of potential Lorentz symmetry 
breaking in the neutrino sector,
to fundamental symmetries and to other conceivable
non-standard interactions~\cite{BaTeTo2019}.
(Note that corresponding questions do not 
occur in Lorentz-symmetry conserving models~\cite{JeEtAl2014}.)
The underlying question is the following:
Is the existence of the LPCR process
compatible with electroweak gauge invariance,
or, do the tight bounds derived in~\cite{StSc2014,StEtAl2015,SoNaJe2019}
(and the theoretical calculations reported
in~\cite{CoGl2011,BeLe2012,SoNaJe2019})
additionally depend on the possibly problematic
assumption of a breaking of $SU(2)_L$
symmetry, in addition to Lorentz symmetry?
These considerations are quite crucial
for the clarification of the status of the 
derived astrophysical bounds on the 
Lorentz-violating parameters~\cite{StSc2014,StEtAl2015,SoNaJe2019}.

We can likewise anticipate that our discussion will lead 
to a fully $SU(2)_L \times U(1)_Y$ gauge invariant model,
with differential Lorentz-symmetry breaking
across generations, that can still accomodate for 
the possibility of neutrino decay, via both NPCR and 
LPCR processes, in the high-energy limit (see Sec.~\ref{sec33}).

This paper is organized as follows.
In Sec.~\ref{sec2}, we introduce a modified Dirac algebra,
adapted to the description of Lorentz-symmetry breaking spin-$1/2$ 
particles. Our investigations continue in Sec.~\ref{sec31} with the 
discussion of a manifestly Lorentz- and gauge-symmetry 
breaking model for the interaction of (conceivable)
superluminal neutrinos with electroweak gauge
bosons; this model has recently been used in 
Refs.~\citen{CoGl2011,BeLe2012}.
In Sec.~\ref{sec32}, we continue with the investigation of 
a model which breaks Lorentz symmetry and (partially)
restores electroweak gauge symmetry, within a 
restricted electroweak  symmetry group.
We continue in Sec.~\ref{sec33} with the 
discussion of a Lorentz-breaking model which fully 
restores the electroweak symmetry group.
Neutrino decay in a fully gauge-invariant model is 
discussed in Sec.~\ref{sec4}.
Conclusions are reserved for Sec.~\ref{sec5}, and
some additional general remarks on the relation of (spontaneous) 
Lorentz-symmetry breaking and gauge invariance are 
relegated to~\ref{appa}.

%
%
\section{Relation to Modified Dirac Algebra}
\label{sec2}

We would like to briefly 
discuss a connection of the 
common Lagrangian used in the description 
of superluminal, Lorentz-violating neutrinos,
to a generalized Dirac algebra.
According to Eq.~(16) of Ref.~\citen{SoNaJe2019},
one may use the following Lagrangian, 
for a Lorentz-violating, massless, left-handed Dirac particle,
\begin{equation}
\calL = \ii \bar\psi \, \gamma_\mu \;
\widetilde g^{\mu\nu}(v) \; \partial_\nu \psi \,,
\end{equation} 
where we assume that $\psi = \nu_\ell^{(m)}$
is a left-handed neutrino mass eigenstate,
i.e., $[(1-\gamma^5)/2] \, \psi = \psi$.
(In the course of the current investigations,
we attempt to keep the notation as concise as possible
and avoid any superfluous superscripts, or
subscripts.) We can write the Lagrangian 
for the massless case as
\begin{equation}
\calL = \bar\psi \, 
\ii \, \widetilde \gamma^\mu \, \partial_\mu \, \psi  \,,
\end{equation}
where the $\widetilde \gamma^\mu$ are given as
\begin{equation}
\label{tildegammas}
\widetilde \gamma^0 = 1, 
\qquad
\widetilde \gamma^i = v \, \gamma^i \,.
\end{equation}
These fulfill the anti-commutator relation
\begin{equation}
\label{example}
\{ \widetilde \gamma^\mu,
\widetilde \gamma^\nu \} =
2 {\widetilde g}^{\mu\nu}(v^2) = 
2 {\rm diag}(1, -v^2, -v^2, -v^2) \,,
\end{equation}
where we note the square of the velocity.
Note that the given pseudo-metric implies a spatially 
isotropic breaking of Lorentz invariance. 
For bounds on coefficients with directional dependence,
we refer to the data compilation presented in Ref.~\citen{KoRu2011}. 

One can in fact relate this formalism to so-called
{\em vierbein} coefficients
(cf.~Refs.~\citen{Je2013,JeNo2013pra,JeNo2014jpa,%
NoJe2015tach,NoJe2016,Je2018geonium}).
Namely, in a more general context,
one can define the generalized Dirac matrices
\begin{equation}
\label{def_vierbein}
{\widetilde \gamma}^\mu = e^\mu_A \, \gamma^A \,,
\end{equation}
where the Einstein summation convention is 
used, and $\gamma^A$ with $A = 0,1,2,3$ 
are the ordinary Dirac $\gamma$ matrices, while
the $e^\mu_A$ take the role of the 
so-called ``vierbein'' in general relativity,
with the property
\begin{equation}
{\widetilde g}^{\mu\nu}(v^2)
= e_\mu^A \, g_{AB} \, e_\nu^B 
= e_\mu^A \, e_{\nu A} \,.
\end{equation}
This implies that 
the ``vierbein'' takes the role of the square root of the 
metric~\cite{JeNo2013pra}.
Capital Latin indices can be raised with the flat-space metric $g^{AB}$.
One can then easily show that
\begin{equation}
\{ {\widetilde \gamma}^\mu, {\widetilde \gamma}^\nu \} =
e^\mu_A \, e^\nu_B \, \{ \gamma^A, \gamma^B \} =
e^\mu_A \, e^\nu_B \, ( 2 g^{A B} ) =
2 \, {\widetilde g}^{\mu\nu}(v^2) \,.
\end{equation}

The analogy to the formalism
of general relativity implies that 
${\widetilde g}^{\mu \nu}(v^2)$ takes the role 
of a modified Lorentz ``metric'', but
without curvature (because we assume that the 
coefficients are constant).
The word ``metric'' should be understood 
with a grain of salt (hence the apostrophes),
because it does not constitute a space-time metric 
in the sense of general relativity,
that is used to measure space-time intervals,
but rather, a mathematical object used to 
parameterize the dispersion relation 
of a Lorentz-violating particle.
One might thus call it a ``pseudo-metric''.
Because of the lack of 
curvature, the pseudo-metric ${\widetilde g}_{\mu \nu}(v^2)$ 
is still characterizing a flat ``space-time''.
For a truly curved space, the 
notation ${\overline g}_{\mu\nu}$ has been 
proposed in Refs.~\citen{Je2013,JeNo2013pra,JeNo2014jpa,%
NoJe2015tach,NoJe2016,Je2018geonium}
in order to distinguish the curved-space quantities
from the flat-space ones.
[As a remark, we here note that the superluminal,
Lorentz-violating neutrino model is 
different from Lorentz-conserving,
$\gamma^5$-Hermitian (``pseudo--Hermitian'') models discussed in the 
literature~\cite{JeWu2012epjc}.]

For a modified ``metric'' of the form~\eqref{example},
one can choose the vierbein coefficients as
\begin{equation}
\label{vier_value}
e^0_0 = 1 \,,
\quad
e^0_i = e^i_0 = 0 \,,
\quad
e^i_j = v \, \delta^i_j \,,
\quad 
i,j = 1,2,3 \,.
\end{equation}
The modified Dirac equation describing the Lorentz
violation, now with a mass term, can be written as
\begin{equation}
\label{LLbreak}
\left( \ii \, {\widetilde \gamma}^\mu \, 
\partial_\mu - m \right) \, \psi = 0 \,.
\end{equation}
We now assume that 
$\psi$ stands for a Majorana neutrino field.
One can multiply from the left by the operator
$(\ii \, {\widetilde \gamma}^\nu \, \partial_\nu + m)$,
and use the operator identity
\begin{equation}
\left( \ii \, {\widetilde \gamma}^\nu \,
\partial_\nu + m \right) \, 
\left( \ii \, {\widetilde \gamma}^\mu \,
\partial_\mu - m \right) 
= -{\widetilde g}^{\mu \nu}(v^2) \,
\partial_\mu \, \partial_\nu - m^2 \,.
\end{equation}
For the metric~\eqref{example}, one can use the 
identity
\begin{equation}
- {\widetilde g}^{\mu\nu}(v^2) \,
\partial_\mu \, \partial_\nu - m^2 = 
E^2 - v^2 \, \vec p^{\,2} - m^2 \,,
\end{equation}
where $E$ is the energy and $\vec p$ is the momentum
operator. This leads to the dispersion relation,
\begin{equation}
\label{disp_rel}
E = \pm \sqrt{\vec p^{\,2} \, v^2 + m^2} \,.
\end{equation}
The Lagrangian~\eqref{LLbreak} is then seen to 
describe a Lorentz-violating particle 
with the dispersion relation~\eqref{disp_rel}.

In order to draw a connection to the
basis of the fermions of the 
Standard Model Extension (SME), 
we refer to the classification
of operators given in Eq.~(9) of Ref.~\citen{KoMe2012}.
Our isotropic Lorentz-violating model corresponds,
in the notation of Eq.~(9) of Ref.~\citen{KoMe2012},
to the case
\begin{equation}
{\hat c}^{\mu\nu}_{F' F"} = \delta_{F' F''} \, c^{\mu\nu} \,,
\qquad
c^{\mu\nu} = 
(v-1) \, ( g^{\mu\nu} - t^\mu \, t^\nu ) \,,
\end{equation}
where the diagonality in the fermion flavor 
indices $F'$ and $F''$ simply means that our Lorentz-violating model
does not involve additional flavor mixing.
The time-like unit vector $t^\mu = (1,0,0,0)$ is 
used throughout this article.)
Note, incidentally, that the 
$\hat\Gamma^\mu_{AB}$ matrices defined in 
Eq.~(9) of Ref.~\citen{KoMe2012} correspond to our 
${\widetilde\gamma}$ matrices, in the sense that 
\begin{equation}
\label{defGamma}
\hat\Gamma^\mu_{F' F''} = \Gamma^\mu \, \delta_{F' F''}  \,,
\qquad
\Gamma^\mu 
= c^{\mu\nu} \, \gamma_\nu = {\widetilde \gamma}^\mu \,.
\end{equation}
Note also that, as pointed out in the text following
Eq.~(9) of Ref.~\citen{KoMe2012}, the 
Lorentz breaking implied by the parameters $c^{\mu\nu}$ 
is CPT even and thus leaves the CPT symmetry intact. 
(It is still interesting to discuss possible
connections of Lorentz symmetry breaking 
and CPT violation; a few remarks on this point
will be given in the following.)
Also, Sec.~II of Ref.~\citen{KoMe2012}
addresses the problem of defining Majorana fermions
in a Lorentz-violating extension of the Standard Model.
Further considerations on this 
point can be found in Ref.~\citen{FuTu2015}.

%
%
\section{Lorentz Violation and Gauge Coupling}
\label{sec3}

%
%
\subsection{Lorentz Violation and Gauge (Non--)Invariance}
\label{sec31}

In this section, we discuss how the 
coupling to the electroweak gauge sector has
to be modified in order to obtain the
effective interaction Lagrangian used
by Cohen and Glashow~\cite{CoGl2011}, which is equivalent 
to ``model I'' used by Bezrukov and Lee~\cite{BeLe2012}.

Let us keep the notation as simple as 
possible, and start from the 
standard generalized Dirac Lagrangian~\eqref{LLbreak},
which we recall for convenience,
\begin{equation}
\calL = \bar\psi (\ii {\widetilde \gamma}^\mu \, \partial_\mu - m ) \psi \,,
\end{equation}
and assume that $\psi$ stands for a 
(Majorana) neutrino field.
(Questions related to the $SU(2)_L$ doublet will 
be answered below.)
We can write this Lagrangian as
\begin{equation}
\label{defcalQ}
\calL = \bar\psi [ \ii \gamma^\mu \partial_\mu 
- m + \underbrace{ \ii ({\widetilde \gamma}^\mu - 
\gamma^\mu ) \partial_\mu }_{\equiv \calQ} ] \psi \,,
\end{equation}
where $\calQ$ is the Lorentz-violating perturbation
and constitutes a special case of Eqs.~(3),~(8) and~(9)
of Ref.~\citen{KoMe2012}.

In Refs.~\citen{KoLe2001,KoMe2009}, 
the $\calQ$ term is advocated to be the sub-Planck limit of 
a nonlocal theory with spontaneous Lorentz and CPT violations~\cite{KoLe2001},
or in more general terms, as the low-energy 
limit of new physics originating from the Planck scale.
Indeed, as we investigate possible violations of
Lorentz invariance, we explore the limits 
of validity of our current understanding of fundamental 
quantum field theory. E.g., it is well known that 
Lorentz invariance is one of the assumptions 
underlying the proof of the CPT theorem~\cite{Jo1957}.
A violation of Lorentz invariance therefore 
allows for violations of CPT, and indeed, some of the 
operators in the full {\em ansatz} for $\calQ$,
as discussed in Ref.~\citen{KoMe2012}, 
are CPT odd. For a long time, one has held
the belief that CPT violation automatically
implies a violation of Lorentz invariance~\cite{Gr2002},
while conversely, broken Lorentz invariance
allows for, but does not require, broken CPT invariance~\cite{Gr2002}.
Recently~\cite{ChDoNoTu2011,ChFuTu2012}, invoking 
additional concepts like nonlocal interactions,
the conclusions of Ref.~\citen{Gr2002} have been questioned,
and it has been claimed that scenarios exist
where CPT invariance is broken, but Lorentz invariance still holds.
In general, the questions regarding the 
ultimate limits of the validity of our current
understanding of fundamental physical laws must 
include bounds on Lorentz-violating terms,
and terms that allow for other broken fundamental
symmetries, like CPT.

Furthermore, the Lorentz-violating operators
are assumed to be the sub-Planck limit of
new physics originating at the Planck scale,
where the fundamental interactions will be 
completely different from ``low-''energy physics
[where ``low-''energy could even extend to the 
PeV scale, which is still twelve orders of 
magnitude below the (reduced) Planck scale
of $\sqrt{1/(8 \pi G)} = 2.4 \times 10^{18} \, {\rm GeV}$].
At ``low'' energy, the coupling to the electroweak 
gauge bosons proceeds by the substitution
\begin{equation}
\label{subst}
\partial_\mu \to D_\mu
\end{equation}
where the operator $D_\mu$ constitutes the $SU(2)_L$ covariant 
derivative, applied to an $SU(2)_L$ doublet, 
as discussed below.
It is therefore permissible, or, suggested,
to experiment with the idea, that the 
substitution~\eqref{subst}
applies {\em only} to the 
unperturbed Lagrangian in Eq.~\eqref{defcalQ},
but leaves the perturbative $\calQ$ term unchanged.
In this case, 
the perturbative term does not participate in the 
electroweak interaction [$SU(2)_L$ doublet], while 
modifying the free propagation of the 
neutrino [once the $\calQ$ operator is written 
so that it applies only to the upper component
of the $SU(2)_L$ doublet, i.e., only to the neutrino]. 

To be specific, let us start from the doublet
\begin{equation}
L_e = \left( \begin{array}{c} \nu_e \\ e_L \end{array} \right)
\end{equation}
[see Eq. (12.227) of Ref.~\citen{ItZu1980}],
where $\nu_e$ is the electron neutrino field,
and $e_L$ is the left-handed electron-positron field,
and consider the coupling to the electroweak
sector, as in Eq.~(12.232) of~\cite{ItZu1980},
concentrating on the terms that couple to the 
electroweak gauge fields [in the Lorentz-covariant theory]
\begin{align}
\label{LG}
\calL_G =& \;
{\barL}_e ( \ii \gamma^\mu D_\mu ) L_e 
\nonumber\\[0.1133ex]
=& \; \barL_e \left[ \ii \gamma^\mu 
\left(\partial_\mu - \frac{g'}{2} B_\mu - g \frac{\tau_i}{2} A_{i,\mu} 
\right) \right] L_e \,,
\end{align}
where the $B$ and the $A_i$ ($i=1,2,3$) fields
transform into the photon, the $Z_0$ and the 
$W^\pm$ gauge bosons under electroweak unification
(for details, see the discussion below).
The Pauli matrices are the $\tau_i$, and they
act within the $SU(2)_L$ doublet.
The charge $e$, and the electroweak couplings 
including the Weinberg angle are related to 
$g$ and $g'$ [see Eq.~\eqref{WeinbergAngle} below].
If we add to $\calL_G$ the mass term
\begin{align}
\label{LM}
\calL_M =& \; - {\barL}_e \cdot {\bf M} \cdot L_e 
\nonumber\\[0.1133ex]
=& \; \left( \begin{array}{c} \bar\nu_e \\ \bar \psi_e \end{array} \right)
\, \left( \begin{array}{cc} - m_\nu + {\calQ} & 0 \\
0 & - m_e \\
\end{array} \right) 
\left( \begin{array}{c} \nu_e \\ \psi_e \end{array} \right),
\end{align}
then the metric to be used for the 
effective interaction (Fermi interaction)
at the electroweak vertex remains the unperturbed
Lorentz metric $g_{\mu\nu}$,
while the propagation of free neutrinos 
acquires a Lorentz-breaking term $\calQ$, 
as specified in Eq.~\eqref{defcalQ}. 
In writing Eq.~\eqref{LM}, we use an oscillation-free
neutrino model, and assume, furthermore,
that the neutrino mass term is of the Majorana type,
i.e., $\nu_e = \nu_e^{(\calC)}$ where $\calC$ denotes
the charge conjugate. We also
supplement the right-handed component of the 
electron field, $\psi_e = e_L + e_R$,
for the Dirac mass term of the electron.
An inspection shows that the Lagrangian
\begin{equation}
\calL = \calL_G + \calL_M 
\end{equation}
directly leads to the interaction Lagrangian 
used by Cohen and Glashow~\cite{CoGl2011} and 
in ``model~I'' of Bezrukov and Lee~\cite{BeLe2012}.
Strictly speaking, the Lagrangian $\calL_G + \calL_M$ 
breaks electroweak gauge invariance 
due to the presence of partial 
(not covariant) derivative operators in $\calQ$,
but the gauge and Lorentz-breaking
terms enter at the same perturbative level,
namely, at first order in $\calQ$ (see also~\ref{appa}).

\subsection{Lorentz--Violation and Gauge Coupling:
One--Flavor Model}
\label{sec32}

In this section, we investigate which 
Lagrangian should be used in the calculation
of vacuum pair emission and neutrino 
splitting if we intend to preserve electroweak
gauge invariance to the extent possible.
We intend to show that it is 
possible to preserve $SU(2)_L$ gauge invariance
under a restricted set of gauge transformations
in the electroweak sector, specifically,
the sector related to the $Z_0$ exchange,
and still break Lorentz invariance differentially,
i.e., with different values for the 
Lorentz-breaking parameters, 
for neutrinos compared to charged fermions.
We first calculate this 
in an ``oscillation-free'' environment (using only one 
particle generation),
where we neglect the mixing 
of neutrino mass eigenstates and weak interaction eigenstates,
due to the off-diagonal entries of the 
Pontecorvo--Maki--Nakagawa--Sakata (PMNS) matrix.

We again emphasize that 
$\tau_i$ matrices in Eq.~\eqref{LG} 
act in the $SU(2)_L$ doublet,
while the $\widetilde\gamma^\mu$ matrices 
act on the electrons and neutrinos separately.
The first observation is that one can choose the 
free Lagrangian as follows
(we ignore the mass terms which are 
irrelevant for the considerations that follow),
\begin{equation}
\label{LF}
\calL_F \sim 
\left( \begin{array}{c} \bar\nu_e \\ \bar e_L \end{array} \right) 
\left( \begin{array}{cc} \ii \widetilde\gamma_{\nu_e}^\mu \partial_\mu & 0 \\
0 & \ii \widetilde\gamma_{e}^\mu \partial_\mu \end{array} \right) \\
\left( \begin{array}{c} \nu_e \\ e_L \end{array} \right) 
\end{equation}
for the $SU(2)_L$ doublet.
In this case, it is immediately clear that
free neutrinos and free electrons obtain different
maximal velocities, according to the 
anti-commutation relations
\begin{subequations}
\begin{align}
\{ \widetilde\gamma_{\nu_e}^\mu, \widetilde\gamma_{\nu_e}^\rho \} = & \;
2 \, {\widetilde g}^{\mu\nu}( v_{\nu_e}^2 ) =
2 \, {\rm diag}(1, -v_{\nu_e}^2, -v_{\nu_e}^2, -v_{\nu_e}^2 \} \,,
\\[0.1133ex]
\{ \widetilde\gamma_{e}^\mu, \widetilde\gamma_e^\rho \} = & \;
2 \, {\widetilde g}^{\mu\nu}( v_e^2 ) =
2 \, {\rm diag}(1, -v_e^2, -v_e^2, -v_e^2 \} \,.
\end{align}
\end{subequations}
As discussed in Sec.~\ref{sec2}, these 
lead to dispersion relations 
$E_e = | \vec p_e | \, v_e$ and 
$E_{\nu_e} = | \vec p_{\nu_e} | \, v_{\nu_e} $
for the electron and electron neutrino, respectively.

The second observation is that one can 
replace the partial derivatives in Eq.~\eqref{LF}
by covariant derivatives, according to Eq.~\eqref{LG}.
The covariant derivation,
under the $SU(2)_L$ gauge group, is matrix-valued
and the substitution $\partial_\mu \to D_\mu$
will lead to off-diagonal entries in Eq.~\eqref{LF},
coupling the electron to the neutrino by what is later identified
as the $W$ boson. Furthermore, the diagonal matrix 
(diagonal with regard to the $SU(2)_L$ doublet) with 
entries
\begin{equation}
\left( \begin{array}{cc}
\widetilde\gamma_{\nu_e}^\mu & 0 \\
0 & \widetilde\gamma_{e}^\mu 
\end{array} \right)
\end{equation}
does not necessarily commute with the $W$ interaction 
Lagrangian, which is proportional to the 
terms involving the $\tau_1$ and $\tau_2$ matrices 
in Eq.~\eqref{LG}. 
However, one can formulate a restricted
set of gauge transformations, which pertain 
only to the 
\begin{equation}
\tau_3 = \left( \begin{array}{cc}
1 & 0 \\
0 & -1 \\
\end{array} \right)
\end{equation}
matrix in Eq.~\eqref{LG}, and restrict the covariant derivative to 
\begin{equation}
\label{defDmu}
\ii \partial_\mu \to \ii D_\mu =
\ii \partial_\mu - \frac{g'}{2} B_\mu + \frac{g}{2} 
\tau_3 \, A_{3,\mu} \,,
\end{equation}
The gauge coupling Lagrangian~$\calL_{Z,A}$ which is 
to be added to $\calL_F$ under the restricted set of 
gauge transformations, reads as follows,
\begin{subequations}
\begin{align}
\calL_{Z,A} = & \bar L_e \cdot {\bf G} \cdot L_e \,,
\\[0.1133ex]
{\bf G} =& \left( \begin{array}{cc}
\frac12 
\widetilde\gamma_{\nu_e}^\mu (g A_{3,\mu} - g' B_\mu ) & 0 \\
0 & -\frac12 
\widetilde\gamma_e^\mu (g A_{3,\mu} + g' B_\mu ) 
\end{array}
\right). 
\end{align}
\end{subequations}
Defining, as in Eq. (12.238) of Ref.~\citen{ItZu1980},
the $Z_0$ and $A_\mu$ fields as
\begin{align}
Z_\mu =& \; \frac{1}{\sqrt{ g^2 + g'^2 }} \, (-g \, A_{3,\mu} + g' B_\mu ) \,,
\\[0.1133ex]
A_\mu =& \; \frac{1}{\sqrt{ g^2 + g'^2 }} \, (g B_\mu + g' A_{3,\mu}) \,,
\end{align}
one obtains the following couplings,
\begin{align}
\label{ZZ}
\calL_{Z,A} =& \; -\frac{e}{2} \left[ \tan\theta_W 
( \bar\nu_e \widetilde\gamma_{\nu_e}^\mu \nu_e + 
\bar e_L \widetilde\gamma_e^\mu e_L ) 
\right. 
\nonumber\\[0.1133ex]
& \; \left. - \cot\theta_W ( \bar e_L \widetilde\gamma_e^\mu e_L - 
\bar\nu_e \widetilde\gamma_{\nu_e}^\mu \nu_e ) \right] \,
Z_\mu 
\nonumber\\[0.1133ex]
& \; - e \, \widetilde\gamma_e^\mu \bar e_L \, A_\mu \, e_L  \,,
\end{align}
where 
\begin{align}
\label{WeinbergAngle}
e =& \; \frac{g g'}{\sqrt{ g^2 + g'^2 }} \,,
\qquad
\tan \theta_W = \frac{g'}{g} \,,
\nonumber\\[0.1133ex]
e =& \; g' \cos\theta_W = g \sin\theta_W  \,,
\end{align}
and $\theta_W$ is the Weinberg angle,
and $e$ is the electron charge.
(Adding the right-handed component of the 
charged fermion field restores the full 
QED Lagrangian for the coupling of the
electron-positron field.)
The result~\eqref{ZZ} is exactly equivalent to the 
corresponding terms in Eq.~(12.240) of~Ref.~\citen{ItZu1980},
with the replacement $\gamma^\mu \to \widetilde\gamma^\mu_{\nu_e}$
for the neutrino couplings to the $Z_0$ boson,
and $\gamma^\mu \to \widetilde\gamma^\mu_e$
for the electron couplings to the $Z_0$ boson.
The resulting modified effective Fermi Lagrangian 
describing the coupling of 
electrons and neutrino 
is thus exactly the one of ``model II'' used by Bezrukov and Lee,
and by us in Ref.~\citen{SoNaJe2019}.
(We recall that the dominant contributions to 
both neutrino as well as lepton-pair
Cerenkov radiation proceed by $Z_0$ exchange.)

For the calculation of neutrino splitting~\cite{SoNaJe2019},
it means that, e.g., if the 
$\widetilde\gamma_{\nu_\mu}^\mu$ for muon neutrinos are 
different from those of electrons neutrinos,
$\widetilde\gamma_{\nu_e}^\mu$, because of a different
maximum velocity for the two species,
then the neutrino splitting process becomes 
kinematically allowed (for $v_{\nu_\mu} = v_f > 
v_{\nu_e} = v_i$).
Furthermore, the effective interaction Lagrangian 
describing the four-fermion vertex
receives a correction from the two $Z_0$ vertices,
leading to the appropriate replacement
\begin{equation}
\label{vivf}
v_{\rm int} = v_i \, v_f 
\end{equation}
for the pseudo-metric to be used in the 
effective Lagrangian in Eq.~(18) of Ref.~\citen{SoNaJe2019},
within a gauge-invariant formulation.
The same is done in ``model II'' of Ref.~\citen{BeLe2012}.
Here, $v_i$ is the Lorentz-violating velocity 
parameter for the initial (oncoming) particle,
while $v_f$ is that of the emitted (final) particle.

In the context of Lorentz-breaking, 
one often finds that symmetry groups 
are broken down to smaller subgroups
(see also the discussion in~\ref{appa}).
Here, we observe that the Lorentz-breaking
terms change the gauge group 
from $SU(2)_L \times U(1)_Y$ to
$U(1)_L \times U(1)_Y$.

\begin{figure} [t]
\begin{center}
\begin{minipage}{0.91\linewidth}
\begin{center}
\includegraphics[width=0.77\linewidth]{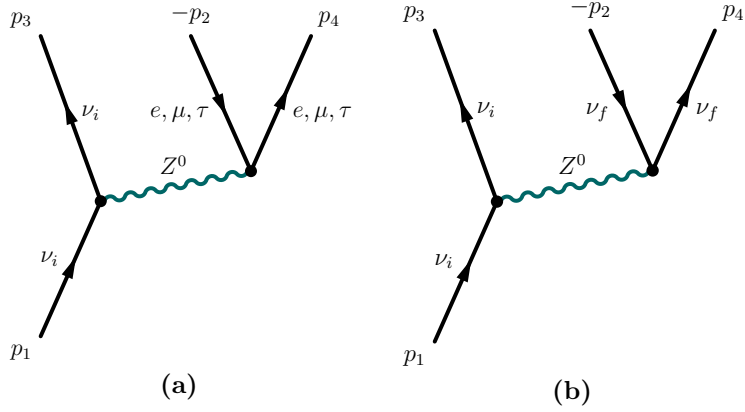}
\caption{\label{fig1}
Feynman diagrams for the LPCR [Fig.~(a)]
and NPCR [Fig.~(b)] processes; these proceed via
exchange of a virtual $Z_0$ boson.
The diagrams are especially relevant in the 
high-energy region of our gauge-invariant, 
but Lorentz-breaking model, where threshold
conditions for LPCR are met. 
The ``faster'' generation of 
particles, of which we assume that $\nu_i$ 
is a member, decays into charged fermions
[Fig.~(a)] or neutal fermions
[Fig.~(b)] of a ``slower'' flavor
(see the text of Sec.~\ref{sec4} for further explanations).
}
\end{center}
\end{minipage}
\end{center}
\end{figure}

\subsection{Lorentz--Violation and Gauge Coupling:
Three--Flavor Model}
\label{sec33}

In the above considerations, we have 
shown that it is possible to formulate 
differential Lorentz violation in the 
same $SU(2)_L$ doublet, to obtain different 
Lorentz-breaking parameters for electron neutrinos 
as compared to (left-handed) electrons,
while preserving gauge invariance with respect
to a restricted subset of $SU(2)_L$ gauge transformations.
This consideration required the use
of different $\widetilde\gamma^\mu$ matrices for the upper
and lower components of the same $SU(2)_L$
doublet.
One might ask the question if different Lorentz-breaking
parameters could be obtained for different
neutrino species, as compared to electrons,
and among the neutrino mass eigenstates,
even if one uses the same
$\widetilde\gamma^\mu$ matrices for the upper
and lower components of the same $SU(2)_L$
doublet, and only assumes a dependence of the 
$\widetilde\gamma^\mu$ matrices on the fermion generations.
In contrast to the model discussed in 
Sec.~\ref{sec2}, we here preserve full
$SU(2)_L$ gauge invariance.

We thus start from the Lagrangian
(ignoring the free mass terms)
\begin{align}
\label{L3G}
\calL_{3G} = & \;
\left( \begin{array}{c} \bar\nu_e \\ \bar e_L \end{array} \right) 
\left( \begin{array}{cc} \ii \widetilde\gamma_{\nu_e}^\mu D_\mu & 0 \\
0 & \ii \widetilde\gamma_{\nu_e}^\mu D_\mu 
\end{array} \right) 
\left( \begin{array}{c} \nu_e \\ e_L \end{array} \right) 
+ (e \leftrightarrow \mu) + (e \leftrightarrow \tau) \,,
\end{align}
where ``$3G$'' refers to the three generations, 
and we assume a uniform Lorentz violation 
within the first generation,
and uniform within the second generation,
but with different overall parameters,
\begin{equation}
\label{differential}
\widetilde\gamma_{\nu_e}^\rho = \widetilde\gamma_e^\rho
\neq
\widetilde\gamma_{\nu_\mu}^\rho = \widetilde\gamma_\mu^\rho 
\neq
\widetilde\gamma_{\nu_\tau}^\rho = \widetilde\gamma_\tau^\rho \,.
\end{equation}
(Note that, in the sense of Sec.~\ref{sec32},
we now have $\widetilde\gamma_{\nu_e}^\mu = \widetilde\gamma_e^\mu$.)
As the matrix 
\begin{equation}
\left( \begin{array}{cc}
\widetilde\gamma_{\nu_e}^\mu & 0 \\
0 & \widetilde\gamma_{\nu_e}^\mu 
\end{array} \right)
\end{equation}
is proportional to the unit matrix [from within the 
$SU(2)_L$ doublet], full gauge invariance is preserved.

Invoking neutrino oscillations, we can write the 
mass term as
\begin{equation}
{\bf M} = \bar \nu_k^{\rm (m)} \, m_k \, \nu_k^{\rm (m)} \,,
\end{equation}
where the $\nu_k^{\rm (m)}$ are the neutrino mass eigenstates
($k = 1,2,3$ is summed over).
In the free theory, we end up with a Lagrangian
\begin{equation}
\label{LFm}
\calL_F = 
\ii \bar \nu_k^{\rm (m)} \; \widetilde \gamma^{{\rm (m)},\mu}_{kj} \;
\partial_\mu  \nu_j^{\rm (m)} - 
\bar \nu_k^{\rm (m)} \, m_k \, \nu_k^{\rm (m)} \,,
\end{equation}
where repeated indices are summed over
the generations ($k,j = 1,2,3$), and 
the mass eigenstates $\nu_k^{\rm (m)}$
and the flavor eigenstates $\nu_j^{\rm (f)}$
are related by the PMNS matrix with entries $U_{kj}$,
\begin{equation}
\nu_k^{\rm (m)} = U_{kj} \, \nu_j^{\rm (f)} \,.
\end{equation}
The emergence of the PMNS matrix 
for both Dirac as well as Majorana neutrinos
is discussed in detail in Ref.~\citen{Bi2015}.
Again, neutrino mass ${\rm (m)}$ and flavor ${\rm (f)}$ eigenstates
are distinguished based on their superscript.
Of course, the mass-basis matrices
\begin{equation}
\widetilde \gamma^{{\rm (m)},\mu}_{kj} =
U_{k\ell} \, \widetilde \gamma^{{\rm (f)},\mu}_\ell \,
U^{-1}_{\ell j}
\end{equation}
are effective, Lorentz-violating, modified Dirac
matrices describing the (possibly off-diagonal, 
$k \neq j$) Lorentz violation in the neutrino
mass eigenstate basis (the subscript $\ell$
is being summed over in the above equation).
[For absolute clarity, we should reemphasize that
the $\widetilde \gamma$ matrices used up to this 
point in our analysis, such as in Eq.~\eqref{L3G},
constitute flavor-basis matrices which would
otherwise carry a superscript $({\mathrm f})$ 
once we distinguish between the mass and the flavor basis.]

Two limiting cases are of interest:
{\em (i)} In the low-energy limit, 
the Lorentz-violating parameters play a 
subordinate role as compared to the 
mass terms, and the energy splitting for equal momenta, among the 
neutrinos, is given in the mass eigenstate 
basis. In that limit, an inspection shows
that the dominant terms in the free
Lagrangian~\eqref{LFm} are just the diagonal 
ones in the mass basis,
\begin{equation}
\label{LFmass}
\calL_F \approx
\ii \bar \nu_k^{\rm (m)} \; \widetilde \gamma^{{\rm (m)},\mu}_{k} \;
\partial_\mu  \nu_k^{\rm (m)} - 
\bar \nu_k^{{\rm (m)}} \, m_k \, \nu_k^{(m)} \,,
\end{equation}
where of course, the subscript $k$ is being summed
over $k=1,2,3$. However, the matrices 
$\widetilde \gamma^{{\rm (m)},\mu}_{\ell}$
are being defined as in 
$\widetilde \gamma^{{\rm (m)},\mu}_{\ell} = 
\widetilde \gamma^{{\rm (m)},\mu}_{\ell\ell}$,
without a summation over~$\ell$. Under these
assumptions, The maximal attainable velocities
$v^{{\rm (m)}}_k$ of the mass eigenstates are thus 
related of the flavor eigenstates, by the relation
\begin{equation}
\sum_{\ell = 1}^3
U_{k \ell} \, v^{\rm (f)}_\ell \, U^{-1}_{\ell k} = v^{\rm (m)}_k \;\;
\mbox{(no summation over $k$)}.
\end{equation}

For the Lorentz-breaking
but gauge-invariant formulation of the 
neutrino splitting process, the appropriate 
choice [for the low-energy region, 
as measured by Eq.~\eqref{criterion}] for the Lorentz-violating parameter
in the effective interaction Lagrangian is
[see Eq.~(18) of Ref.~\citen{SoNaJe2019}]
\begin{equation}
\label{vint}
v_{\rm int} = v_i^{\rm (m)} \, v_f^{\rm (m)}
\end{equation}
where the velocities of the initial and final mass
eigenstates are denoted as $v_i^{\rm (m)}$ and $v_f^{\rm (m)}$, respectively.
Furthermore, it is clear that the effective velocities
$v_k^{(m)}$ for the neutrino mass eigenstates, under the given 
assumptions, will be different from those of the 
electrons, which are given (due to the absence 
of mass mixing among the {\em charged} leptons)
by $v_j^{(f)}$, thus kinematically allowing the vacuum-pair emission
process [again, for the low-energy region,
as measured by Eq.~\eqref{criterion}].

{\em (ii)} In the high-energy limit, one 
can neglect the mass term in Eq.~\eqref{LFm},
and observes that in this limit, 
the flavor eigenstates approximate the mass eigenstates.
Furthermore,
the PMNS matrix approaches the unit matrix,
\begin{equation}
U_{k \ell} \to \delta_{k \ell}
\qquad
\mbox{(high-energy limit with $v_e \neq v_\mu \neq v_\tau$).}
\end{equation}
Here, we refer to Eq.~\eqref{differential} for the definition of 
the corresponding ${\widetilde \gamma}$ matrices, with
\begin{align}
\bigl\{ {\widetilde\gamma_{\nu_e}^\rho, \widetilde\gamma_{\nu_e}^\sigma} \bigr\} = & \;
\bigl\{ \widetilde\gamma_{e}^\rho, \widetilde\gamma_{e}^\sigma \bigr\} =
2 {\widetilde g}^{\rho\sigma}(v_e^2) \,,
\\[0.1133ex]
\bigl\{ {\widetilde\gamma_{\nu_\mu}^\rho, \widetilde\gamma_{\nu_\mu}^\sigma} \bigr\} = & \;
\bigl\{ \widetilde\gamma_{\mu}^\rho, \widetilde\gamma_{\mu}^\sigma \bigr\} =
2 {\widetilde g}^{\rho\sigma}(v_\mu^2) \,,
\\[0.1133ex]
\bigl\{ {\widetilde\gamma_{\nu_\tau}^\rho, \widetilde\gamma_{\nu_\tau}^\sigma} \bigr\} = & \;
\bigl\{ \widetilde\gamma_{\tau}^\rho, \widetilde\gamma_{\tau}^\sigma \bigr\} =
2 {\widetilde g}^{\rho\sigma}(v_\tau^2) \,,
\end{align}
where the subscript $e$, $\mu$ and $\tau$ refer to the different
fermion flavors (generations).
For absolute clarity, we reiterate that, according to the discussion above, the 
velocities $v_e$, $v_\mu$ and $v_\tau$ are defined, first and foremost,
in the flavor eigenstate basis, with the charged fermions and neutrinos
within the same generation attaining the same velocity.

The Lagrangian, in the high-energy limit, can be written as
\begin{equation}
\label{LFflavor}
\begin{split}
& \calL_F \approx 
\ii \bar \nu_k^{\rm (f)} \; \widetilde \gamma^{{\rm (f)},\mu}_k \;
\partial_\mu  \nu_k^{\rm (f)} \approx
\ii \bar \nu_k^{(m)} \; \widetilde \gamma^{(m),\mu}_k \;
\partial_\mu  \nu_k^{(m)} \,,
\\[0.1133ex]
& \gamma^{(m),\mu}_k \to
\gamma^{(f),\mu}_k \,,
\qquad
\nu_k^{(m)} \to
\nu_k^{(f)} \,,
\qquad
\mbox{(high-energy limit with $v_e \neq v_\mu \neq v_\tau$)} \,,
\end{split}
\end{equation}
where we remember that we started from 
the $\widetilde \gamma^{{\rm (f)},\mu}_k$ matrices which were diagonal 
in the flavor basis [see Eq.~\eqref{L3G},
with $k = e,\mu\,\tau$].
Under this assumption, both vacuum pair emission 
(Refs.~\citen{CoGl2011,BeLe2012}) 
as well as neutrino splitting are kinematically 
allowed across (but not within!) generations (flavors), 
provided we have
\begin{equation}
\label{faster}
v_f < v_i \qquad
(f,i = e,\mu,\tau) \,.
\end{equation}
That is to say, the ``faster generation''
decays into the ``slower generation''.
We recall once more that, within the gauge-invariant model,
the charged fermions offer the same
Lorentz-violating parameters as the 
corresponding neutrino flavors,
and hence, $v_f \equiv v^{\rm (f)}_k$.
Neutrino splitting
as well as vacuum pair emission are both
kinematically allowed,
because of the differences among the Lorentz-violating
parameters for the different neutrino flavors,
which happen to approximate the mass (energy)
eigenstates under the given assumptions.

The coincidence of the mass and flavor eigenstates
in the high-energy limit makes the theoretical analysis 
easier; the appropriate choice for the 
parameter $v_{\rm int}$ (see Ref.~\citen{SoNaJe2019}) 
entering the interaction Lagrangian 
(see also Sec.~\ref{sec4} below) is
\begin{equation}
\label{vintf}
v_{\rm int} = v_i \, v_f = v_i^{\rm (f)} \, v_f^{\rm (f)} 
\qquad
\mbox{(high-energy limit with $v_e \neq v_\mu \neq v_\tau$)} \,,
\end{equation}
and it fully preserves preserves gauge invariance.

The transition among the two regimes 
characterized by the Lagrangians~\eqref{LFflavor}
and~\eqref{LFmass} occurs at a momentum scale 
of the order of
\begin{equation}
\label{criterion}
|\vec p| = \sqrt{ \delta m^2/ \delta_{f_1 f_2} } \,,
\end{equation}
where $\delta m^2$ is a typical neutrino mass square difference, and of 
course, $\delta_{f_1 f_2}$ is a typical delta-parameter difference among 
the Lorentz-violating parameters for the different neutrino flavors
(We set $v^2 = 1 + \delta$, in accordance with
Refs.~\citen{CoGl2011,BeLe2012,SoNaJe2019}.)
For a parameter estimates of two different neutrino flavors
of $\delta_{f_1 \, f_2} \sim 10^{-20}$ and 
$\delta m^2 \sim 10^{-3} \, {\rm eV}^2$, 
the transition should occur at momenta 
on the order of $10^8 \dots 10^9 \, {\rm eV}$.
(We here refer to bounds on Lorentz-violating
parameters from laboratory-based experiments~\cite{CoGl1999,AAEtAl2013boone},
which are less strict than those derived
from astrophysical observations~\cite{AaEtAl2010lorentz,AaEtAl2018lorentz};
the latter, though, are under less stringent external control.)

%
%
\section{Gauge Invariance and Neutrino Decay}
\label{sec4}

The observations made above, especially 
those reported in Sec.~\ref{sec33} for the high-energy 
limit of differential Lorentz violation across 
generations (flavors), but with the same Lorentz-violating
parameters ascribed to charged and neutral fermions,
allow us to discuss a fully gauge-symmetry conserving 
model, which still allows for NPCR and LPCR decays
to proceed. Compared with other models 
studied in the literature (see Refs.~\citen{CoGl2011,BeLe2012,SoNaJe2019}), 
the model discussed here
is most restricted in parameter space (it requires
flavor-dependent Lorentz-breaking parameters),
but perhaps, most stringent in its theoretical 
formulation, in the sense that it can be fully embedded
into the Lorentz-violating Standard Model Extension
(SME). Because of full gauge invariance, 
we are also able to address, in passing, the 
gauge dependence not within the $SU(2)_L \times U(1)_Y$ gauge group,
but within additional terms induced by 
the $R_\xi$ gauge for the $Z^0$ boson propagator, 
mapped onto the effective Fermi interaction.
Another question to address is whether 
the $SU(2)_L$ gauge symmetry protects potentially
superluminal neutrinos from the NPCR and LPCR decay processes
discussed in Refs.~\citen{CoGl2011,BeLe2012,SoNaJe2019}.
Or, one might ask if, conversely, bounds on
Lorentz-violating parameters for charged
fermions could universally be applied to 
neutrinos, if we postulate the full retention
of $SU(2)_L \times U(1)_Y$ gauge invariance.

Under the assumption of flavor-dependent 
Lorentz breaking, if, say, electrons and electron
neutrinos propagate faster than muons and muon neutrinos,
the decay processes
$\nu_e \to \nu_e \, \nu_\mu \bar\nu_\mu$, and 
$\nu_e \to \nu_e \, \mu^- \, \mu^+$, 
will be kinematically allowed [see Eq.~\eqref{faster}].
Let us give a brief account of the calculations. 
We define the pseudo-metric corresponding to 
the velocity $v_j > 1$, in full accordance with 
Eq.~\eqref{example}, as follows,
\begin{equation}
{\widetilde g}^{\mu\nu}(v_j) = 
v_j \, g^{\mu\nu} + (1-v_j) t^\mu t^\nu = 
{\rm diag}(1, -v_j, -v_j, -v_j) \,.
\end{equation}
The massive Feynman propagator for the 
gauge vector boson  in $R_\xi$ gauge is
given as follows,
\begin{equation}
D_F^{\mu\nu} = - \frac{g^{\mu\nu} + (\xi -1 ) k^\mu k^\nu/(k^2 - \xi M_Z^2) }%
{k^2 - M_Z^2 + \ii \epsilon} \,,
\end{equation}
where $\epsilon > 0$ denotes the infinitesimal imaginary part,
and $M_Z$ is the $Z_0$ boson mass.
The modified Dirac matrices ${\widetilde \gamma}^\mu_j$,
which are alternatively denoted as $\Gamma^\mu_j$
[see Eq.~\eqref{defGamma}], read as follows,
\begin{equation}
\Gamma^\mu_j = 
{\widetilde \gamma}^\mu_j = 
\left[ v_j \, g^{\mu\nu} + (1-v_j) \, t^\mu \, t^\nu \right] = 
{\widetilde g}^{\mu\nu}(v_j) \, \gamma_\nu \,,
\end{equation}
where we note that there
is no distinction any more between the
flavor and the mass eigenstate basis.
The effective Lagrangian for the 
decay $\nu \to \nu \Psi {\overline \Psi}$ is given as follows,
\begin{align}
\calL_{\rm int}^{2\nu 2\Psi} = & \;
\frac{G_F}{2 \sqrt{2}} \,
\left[ {\overline \nu}_j \, {\widetilde \gamma}^\mu_j
\, (1 - \gamma_5) \, \nu_j \right] \,
g_{\mu\nu} \,
\left[ {\overline \Psi}_k \, {\widetilde \gamma}^\nu_j
\, (c_V  - c_A \, \gamma_5) \, \Psi_j \right] 
\nonumber\\[0.1133ex]
= & \;
\frac{G_F}{2 \sqrt{2}} \,
\left[ {\overline \nu}_j \, {\widetilde \gamma}^\mu_j 
\, (1 - \gamma_5) \, \nu_j \right] \,
{\widetilde g}_{\mu\nu}(v_j \, v_k) \,
\left[ {\overline \Psi}_k \, {\widetilde \gamma}^\nu_j
\, (c_V - c_A \, \gamma_5) \, \Psi_j \right] \,.
\end{align}

For NPCR, we have
\begin{equation}
\Psi_k = \nu_k 
\qquad
\Rightarrow
\qquad
c_V = c_A = 1 \,,
\end{equation}
whereas if $\ell_k$ is a charged lepton (electron--positron pair), 
then
\begin{equation}
\Psi_k = \ell_k 
\qquad
\Rightarrow
\qquad
c_V = 0 \,,
\qquad
c_A = -1/2 \,,
\end{equation}
approximately.
The invariant matrix element in the full gauge theory is
\begin{align}
\calM =& \; \frac{g^2}{16 \, \cos^2 \theta_W} \,
\left[ {\overline u}_i(p_3) \, {\widetilde \gamma}_i^\mu \, 
(1 - \gamma^5) u_i(p_1) \right] \,
\frac{\ii}{(p_2 + p_4)^2 - M_Z^2 + \ii \epsilon} 
\nonumber\\[0.1133ex]
& \; \times \left[ g_{\mu\nu} + (\xi - 1) \, 
\frac{(p_2 + p_4)_\mu \, (p_2 + p_4)_\nu}%
{(p_2 + p_4)^2 - \xi M_Z^2} \right] \,
\left[ {\overline u}_f(p_4) \, {\widetilde \gamma}_f^\nu \, 
(c_V - c_A \, \gamma^5 ) u_f(p_2) \right] \,,
\end{align}
where, just like in Ref.~\citen{SoNaJe2019},
$p_1$ is the four-momentum of the oncoming particle,
$p_3$ is the neutrino momentum after the decay,
and $p_2$ and $p_4$ are the four-momenta of the 
created fermion--anti-fermion pair (see Fig.~\ref{fig1}).
The squared matrix element computed in this
way is gauge invariant (with respect to the electroweak gauge group), 
and an explicit calculation shows
that terms involving the gauge parameter $\xi$ of the $R_\xi$ gauge 
do not appear
in the final result. In order to arrive at this result, it is
crucial though that
{\em (i)} 
the $Z_0$ boson--fermion--€"anti-fermion vertices are proportional to ${\widetilde \gamma}_j^\mu$,
{\em (ii)} the correct prescription for the spin sums
[see Eq.~(32) of Ref.~\citen{SoNaJe2019}] is used, 
\begin{align}
\label{eq:spinsum-SL-t}
\sum_s \, \nu_{j,s} \otimes \overline \nu_{j,s} = & \;
{\widetilde g}_{\mu\nu}(v_j) \, \gamma^\mu \, p^\nu
= v_i \slashed{p} + (1-v_i) (p\cdot t)\slashed{t} \,,
\end{align}
{\em (iii)}~and that the dispersion relation in Eq. (22) is taken into 
account for external superluminal particles, most conveniently in the form 
$v_j p^2 + (1 - v_j)^2 (t \cdot p)^2 = 0$.

We note that in a spontaneously broken gauge theory,
the gauge invariance of the squared matrix element 
computed in $R_\xi$ gauge is usually only recovered once 
diagrams involving both vector and scalar boson exchanges
are summed up. (This may involve exchanges of 
gauge bosons of the weak interaction as well as
Higgs particle exchanges.) Indeed, in the Standard Model, the
left-handed lepton current ${\overline \psi} \gamma_\mu 
(c_V - c_A \gamma^5 ) \psi$ is not conserved, 
with the non-conservation being proportional to
the fermion mass. (Note that this non-conservation
already occurs at tree level and can be derived 
on the basis of the axial component of the current.) 
However, the mass itself is proportional to the 
Yukawa coupling of the fermion (to the Higgs particle). 
Since we are working in the massless 
approximation for leptons,
the Yukawa couplings in our model are set to zero, and
the single diagram with only $Z_0$ boson exchange is gauge
invariant by itself. In particular, we have 
explicitly checked that terms coming from contractions involving the
part of the $Z_0$ boson propagator proportional to 
$(\xi - 1)$ sum to zero upon using the superluminal 
dispersion relation.

For a general decay process in our 
gauge-invariant model, we obtain the 
following formula for the decay rate,
\begin{align}
\label{eq:Gamma-general}
\Gamma_{\nu_i \to \nu_i \psi_f \overline{\psi}_f} =& \; 
\left( \frac{g^2}{16 \, M_Z^2 \, \cos^2 \theta_W } \right)^2 \,
\frac{E_1^5}{24 \pi^3} \,
\frac{c_V^2 + c_A^2}{420 \, n_s} \,
\nonumber\\[0.1133ex]
& \; \times (\delta_i - \delta_f) \, 
\left[ 17 (\delta_i - \delta_f)^2 + 7 \, (\delta_i + \delta_f)^2 \right] \,.
\end{align}
Here, $n_s$ counts the allowed spin states of the neutrino
($n_s=2$ in Ref.~\citen{CoGl2011} but $n_s = 1$
in Ref.~\citen{BeLe2012}).
Note that the factor $f_e$ used in the notation 
of Ref.~\citen{SoNaJe2019}
is absorbed in the prefactor $c_V^2 + c_A^2$.
The energy loss rate is obtained, in the fully 
gauge-invariant model, as
\begin{align}
\label{eq:dEdx-general}
\frac{\dd E_{\nu_i \to \nu_i \psi_f \overline{\psi}_f}}{\dd x} 
=& \; \left( \frac{g^2}{16 \, M_Z^2 \, \cos^2 \theta_W } \right)^2 \,
\frac{E_1^6}{24 \pi^3} \,
\frac{c_V^2+c_A^2}{672 n_s} 
\nonumber\\[0.1133ex]
& \; \times (\delta_i-\delta_f) 
\left[ 22 (\delta_i - \delta_f)^2 + 8 \, (\delta_i + \delta_f)^2 \right] \,.
\end{align}

For LPCR decay, in our gauge-invariant (GI) model, we find
\begin{subequations}
\begin{align}
\label{resLPCR}
\Gamma_{\nu_i \to \nu_i e^- e^+} &= a_{\mathrm{GI}} \frac{G_F^2}{192\pi^3} k_1^5\,,
\\
\frac{\dd E_{\nu_i \to \nu_i e^- e^+}}{\dd x} &= -a'_{\mathrm{GI}} \frac{G_F^2}{192\pi^3} k_1^6\,,
\end{align}
\end{subequations}
with the following results,
\begin{subequations}
\label{resa}
\begin{align}
a_{\mathrm{GI}} =& \; 
\frac{17 (c_V^2 + c_A^2)}{420 n_s} 
(\delta_i - \delta_f) \, \left[ (\delta_i - \delta_f)^2 
+ \frac{7}{17} \, (\delta_i + \delta_f)^2 \right] \,, 
\\
a'_{\mathrm{GI}} =& \;
\frac{11 (c_V^2 + c_A^2)}{336 n_s} 
(\delta_i - \delta_f) \, \left[ (\delta_i - \delta_f)^2 
+ \frac{4}{11} \, (\delta_i + \delta_f)^2 \right] \,.
\end{align}
\end{subequations}
For NPCR decay, in our gauge-invariant model, we find
\begin{subequations}
\begin{align}
\label{resNPCR}
\Gamma_{\nu_i \to \nu_i \nu_f \overline \nu_f} &= 
b_{\mathrm{GI}} \frac{G_F^2}{192\pi^3} k_1^5\,,
\\
\frac{\dd E_{\nu_i \to \nu_i \nu_f \overline \nu_f}}{\dd x} 
&= -b'_{\mathrm{GI}} \frac{G_F^2}{192\pi^3} k_1^6\,,
\end{align}
\end{subequations}
where the coefficients 
$b_{\mathrm{GI}}$ and $b'_{\mathrm{GI}}$ 
can be obtained from 
Eq.~\eqref{eq:Gamma-general}
by setting 
$c_V = c_A = 1$,
\begin{subequations}
\label{resb}
\begin{align}
b_{\mathrm{GI}} =& \;
\frac{17}{210 n_s}
(\delta_i - \delta_f) \, \left[ (\delta_i - \delta_f)^2
+ \frac{7}{17} \, (\delta_i + \delta_f)^2 \right] \,,
\\
b'_{\mathrm{GI}} =& \;
\frac{11}{168 n_s}
(\delta_i - \delta_f) \, \left[ (\delta_i - \delta_f)^2
+ \frac{4}{11} \, (\delta_i + \delta_f)^2 \right] \,.
\end{align}
\end{subequations}
The change in the prefactors as compared to the 
kinematics-only approach pursued in Refs.~\citen{CoGl2011,BeLe2012,SoNaJe2019}
does not significantly change the conclusions drawn in 
Ref.~\citen{SoNaJe2019} on astrophysically 
derived bounds for the Lorentz-violating parameters.
We can establish that $SU(2)L \times U(1)_Y$ gauge
invariance does {\em not} protect superluminal neutrinos
from decay and energy loss processes (NPCR and LPCR). 

%
%
\section{Conclusions}
\label{sec5}

In this paper, we have investigated the assumptions
underlying the model dependent interaction Lagrangians
used in~\cite{CoGl2011,BeLe2012,SoNaJe2019} 
for the formulation of the lepton-pair (LPCR) and neutrino-pair 
Cerenkov radiation (NPCR, see Sec.~\ref{sec4}) processes, 
which have led to very tight
bounds on the Lorentz-violating parameters
in the neutrino sector~\cite{StSc2014,StEtAl2015,SoNaJe2019}. 
The main results can be summarized as follows.
 
{\em Conclusion (i).} 
The model used by Cohen and Glashow~\cite{CoGl2011}, 
and ``model I'' of Bezrukov and Lee~\cite{BeLe2012}
can be traced to an interaction 
Lagrangian which breaks electroweak gauge
invariance, in addition to Lorentz invariance
(see Sec.~\ref{sec31}).
However, this breaking proceeds on the same perturbative
level on which the Lorentz-breaking terms themselves
are formulated [see Eq.~\eqref{LM}].
A discussion on the implications with respect
to fundamental symmetries is given in Sec.~\ref{sec31}
of this article.
 
{\em Conclusion (ii).} ``Model II'' of Bezrukov and Lee, used in the 
formulation of the LPCR process in Ref.~\citen{BeLe2012}, 
and also used by us in Ref.~\citen{SoNaJe2019}, 
is gauge invariant under a restricted set
of gauge transformations, within the 
$SU(2)_L$ gauge group.
The use of non-uniform modified Dirac 
matrices, within the same $SU(2)_L$ doublet, 
is crucial to this observation [see Eq.~\eqref{LF}].
The derivation goes through even in
an ``oscillation-free'' environment where one neglects
the off-diagonal entries of the PMNS matrix,
in the neutrino sector.
The result given in Eq.~\eqref{vivf} clarifies 
the ``gauge-invariant'' Lagrangian
used in ``model II'' [see Eq.~(4) of Ref.~\citen{BeLe2012}].
 
{\em Conclusion (iii).} If one invokes neutrino oscillations,
then the situation is even more favorable for the 
gauge-invariant models (see Sec.~\ref{sec33}).
One can use uniform modified Dirac matrices within the 
same $SU(2)_L$ doublet, but assumes
different Lorentz-violating
parameters between generations [see Eq.~\eqref{L3G}].
By assuming only a generation dependence,
one obtains differential Lorentz violation 
among the neutrino mass eigenstates,
and between neutrinos and charged leptons,
without breaking $SU(2)_L \times U(1)_Y$ gauge invariance.
Under these assumptions [see Eq.~\eqref{vint}], it is useful
to keep the Lorentz-violating parameters
$v_{\rm int}$ that enters the interaction Lagrangian,
separate from the ones of the initial and final states,
as is done in Ref.~\citen{SoNaJe2019}.
Inspired by the considerations reported in Sec.~\ref{sec33},
one may devise a fully $SU(2)_L \times U(1)_Y$ gauge-symmetry 
conserving model, which still allows 
for the NPCR and LPCR decays to proceed (see Sec.~\ref{sec4}).
In the course of the calculations reported in Sec.~\ref{sec4},
we also address the question regarding the dependence of 
the results on the gauge used for the massive vector boson propagator
($R_\xi$ gauge).
 
We have thus clarified the 
cryptic remark of the ``gauge invariance''
of ``model II'' of Bezrukov and Lee
[see Eq.~(4) of Ref.~\citen{BeLe2012}],
and provided additional motivation 
for the functional form of the 
various model-dependent interaction Lagrangians
used in Refs.~\citen{CoGl2011,BeLe2012,SoNaJe2019}.

The very stringent bounds on the Lorentz-violating
parameters in the neutrino sector, based 
on astrophysical observations~\cite{StSc2014,StEtAl2015,SoNaJe2019},
thus do not require models in which electroweak gauge
invariance is broken.
This observation is quite crucial because it implies that 
one cannot ``argue away'' the tight bounds 
derived in Refs.~\citen{StSc2014,StEtAl2015,SoNaJe2019} for the Lorentz-breaking 
parameters in the neutrino sector,
based on the notion that the 
preservation of electroweak gauge invariance
would otherwise preclude the existence of the 
decay processes on which the bounds are based.

%
%
\section*{Acknowledgements}

The authors would like to acknowledge
support from the National Science Foundation (Grant
PHY--1710856). 
This work was also supported by the \'UNKP-17-3 New National Excellence
Program of the Ministry of Human Capacities of Hungary
and by grant K 125105 of the National Research, Development
and Innovation Fund in Hungary.

\appendix

%
%
\section{Spontaneous Lorentz--Symmetry Breaking:
Models and Implications}
\label{appa}

Although the {\em ansatz} of the current paper is completely
phenomenological, and we do not discuss the 
possible mechanism behind Lorentz violation 
in any greater detail, it is still instructive
to mention a specific model of spontaneous 
Lorentz invariance violation,
which has been discussed in rather great detail 
in the literature.

Namely, according to Refs.~\citen{He1957,Bj1963,Eg1976},
the photon could potentially be formulated as 
the Nambu-Goldstone boson linked to spontaneous Lorentz invariance violation.
(This {\em ansatz} was originally formulated before 
electroweak unification.)
Interest in this approach has recently been revived, 
and the theory has been worked out in greater detail~\cite{He1957,Bj1963,BB1963,Eg1976,%
ChFrNi2001prl,ChFrNi2001npb,Bj2001,AzCh2006,ChJe2007,ChFrJeNi2008,ChJe2008,ChFrNi2009}.
Both  Abelian as well as a non-Abelian gauge
theories have been discussed~\cite{ChFrNi2001prl}.
In the case of an Abelian gauge theory,
one assumes that the gauge field $A_\mu$
obtains a non-vanishing vacuum expectation value
according to
[see text after Eq.~(1) of Ref.~\citen{ChJe2008}],
\begin{equation}
\langle A_\mu \rangle = n_\mu \, M \,,
\end{equation}
where $M$ is a (possibly large) energy scale
at which the breaking of Lorentz symmetry occurs.
The Lorentz group restricts itself to $SO(1,2)$ if $n_\mu$
is space-like ($n_\mu \, n^\mu = -1$), 
and into $SO(3)$ if $n_\mu$ is timelike ($n_\mu n^\mu = 1$).

The dynamical constraint
[see Eq.~(1) of Ref.~\citen{ChJe2008}]
\begin{equation}
A_\mu \, A^\mu = n^2 \, M^2 \,,
\end{equation}
is imposed on the $A_\mu$ field.
One then parameterizes the $A_\mu$
field as [see Eq.~(3) of  Ref.~\citen{ChJe2008}]
\begin{equation}
A_\mu = a_\mu + \frac{n_\mu}{n^2} ( n \cdot A ) \,,
\end{equation}
where the $a_\mu$ takes the role of the photon field.
The following Lagrangian is eventually obtained
after an expansion in leading order in $1/M$
[see Eq.~(3) of Ref.~\citen{ChJe2008}],
\begin{align}
\label{israel_lancho}
L(a, \psi) = & \;
-\frac14 \, f_{\mu\nu} \, f^{\mu\nu} -
\frac12 \, \delta  (n \cdot a)^2
\nonumber\\[0.1133ex]
& \;
+ \bar\psi (\ii \gamma^\mu \partial_\mu - m) \psi
- e a_\mu \, \bar\psi \, \gamma^\mu \, \psi
\nonumber\\[0.1133ex]
& \; 
-\frac14 \, f_{\mu\nu} \, h^{\mu\nu} \, 
\frac{n^2 a_\rho \, a^\rho}{M}
+ \frac{e n^2 a_\rho \, a^\rho}{2 M} 
\bar\psi (\gamma \cdot n) \psi \,.
\end{align}
Here, $a_\mu$ takes the role 
of the (quantized) electromagnetic field,
while $f_{\mu\nu} = \partial_\mu a_\nu - \partial_\nu a_\mu$
is the field strength tensor.
Also, $h^{\mu\nu} = n^\mu \partial^\nu - n^\nu \partial^\mu$
is an oriented Lorentz-violating tensor.
The orthogonality condition $n \cdot a = 0$
is explicitly introduced in the Lagrangian 
through a gauge-fixing term with parameter $\delta$.
Note that the Lagrangian~\eqref{israel_lancho}
is obtained after a suitable redefinition
of the fermion field, as given explicitly in Eq.~(6) 
of Ref.~\citen{ChJe2008}.

We note that the sum of the terms
\begin{equation}
\bar\psi (\ii \gamma^\mu \partial_\mu - m) \psi
- e a_\mu \, \bar\psi \, \gamma^\mu \, \psi
\end{equation}
in Eq.~\eqref{israel_lancho} add up to the 
gauge-invariant quantum electrodynamic 
interaction ($e$ is the electron charge).

In the limit $M \to \infty$, 
the entire Lagrangian~\eqref{israel_lancho}
approximates the ordinary QED Lagrangian.
However, for finite $M$,
the fifth and the sixth term on the right-hand
side of Eq.~\eqref{israel_lancho},
which are initially generated by spontaneous
Lorentz breaking in the electromagnetic sector, 
explicitly break electromagnetic gauge invariance,
in addition to breaking Lorentz invariance.
The fifth term generates a three-photon
vertex, while the sixth term generates a 
two-fermion, two-photon interaction (see, e.g.,
Ref.~\citen{AzCh2006}).

In Eq.~(22) of Ref.~\citen{AzCh2006},
it is shown that the contribution of both of the 
Lorentz-breaking terms to the electron-photon
scattering amplitude vanishes (due to mutual cancelations) 
if we take the matrix element between on-shell spinors.
Around Eq. (32) of Ref.~\citen{AzCh2006}, it is 
argued that the same cancelation occurs 
for the one-loop amplitude,
if the specific photon propagator integral
given in Eq.~(32) of Ref.~\citen{AzCh2006}
is evaluated in dimensional regularization.
These considerations show that the 
Lorentz-violating terms in Eq.~\eqref{israel_lancho}
do not necessarily lead to observable
effects at low-energy.

The generalization to spontaneous 
Lorentz-symmetry breaking in non-Abelian 
gauge fields involves the assumption
[see Eq.~(9) of Ref.~\citen{ChFrNi2001prl}]
\begin{equation}
\langle A^i_\mu \rangle = n^i_\mu \, M
\end{equation}
where the upper index $i$
describes the component within the non-Abelian
gauge group, e.g., $SU(N)$,
in which case $i=1,\dots,N$.
For the Lorentz-breaking terms to vanish in the low-energy 
limit, one then has to make additional assumptions
regarding the 
masses of the particles in a given $SU(N)$
multiplet; e.g., according to 
Eq.~(19) of Ref.~\citen{ChFrNi2001prl},
one needs to assume these masses to be equal.

For the context of the current paper, 
two observations are relevant:

{\em (i)} The approach taken in 
Refs.~\citen{ChFrNi2001prl,ChFrNi2001npb,Bj2001,AzCh2006,%
ChJe2007,ChFrJeNi2008,ChJe2008,ChFrNi2009}
starts from a spontaneous Lorentz symmetry 
breaking at some high-energy scale $M$,
involving a gauge boson field.
This assumption is quite natural, because
symmetry breaking for a vector field 
automatically singles out a specific direction in 
space-time (it would not necessarily do so for a spinor). 
However, as a comparison to 
Eq.~\eqref{israel_lancho} shows, for the case of
spontaneous symmetry breaking in the gauge boson sector,
the fermion sector is largely 
unaffected by the Lorentz-symmetry breaking,
which initially occurs only in the gauge
boson sector. [We observe that the third and fourth term
in Eq.~\eqref{israel_lancho} add up to the 
fully Lorentz-covariant Lagrangian for the 
electromagnetically coupled electron.] 
Hence, the {\em ansatz} discussed
in Refs.~\citen{He1957,Bj1963,BB1963,Eg1976,%
ChFrNi2001prl,ChFrNi2001npb,Bj2001,AzCh2006,ChJe2007,ChFrJeNi2008,ChJe2008,ChFrNi2009}
is not directly applicable to the models constrained by our 
calculations, which pertain to Lorentz violation in the
fermion (neutrino) sector.

{\em (ii)} A very important 
observation can be made. Namely,
Lorentz violation and gauge invariance 
violation are intimately intertwined.
The term
\begin{equation}
\label{term}
\frac{e n^2 a_\rho \, a^\rho}{2 M} 
\bar\psi (\gamma \cdot n) \psi
\end{equation}
in Eq.~\eqref{israel_lancho}
manifestly breaks the electromagnetic $U(1)_{\rm EM}$ gauge 
symmetry. We remember that a gauge transformation 
in quantum electrodynamics works as 
$a_\mu \to a_\mu - \partial_\mu \Lambda$
and $\psi \to \psi \, \exp(\ii e \Lambda)$,
where $\Lambda = \Lambda(x)$ is the gauge function,
and $e$ is the electron charge.
Under this gauge transformation, the 
term~\eqref{term} is manifestly non-invariant.
Lorentz violation has thus 
created a term that violates gauge invariance, on the 
perturbative level (in first order in the $1/M$ 
expansion). Analogously, the model used by Cohen and Glashow
in Ref.~\citen{CoGl2011} assumes a breaking of 
gauge invariance on the perturbative level,
in the latter case, of the electroweak gauge symmetry.
Based on our comparison with the 
approach taken in 
Refs.~\citen{He1957,Bj1963,BB1963,Eg1976,%
ChFrNi2001prl,ChFrNi2001npb,Bj2001,AzCh2006,ChJe2007,ChFrJeNi2008,ChJe2008,ChFrNi2009},
this is a perfectly permissible assumption.

\end{document}